%% file: draftv1.tex
\documentclass[aps,prl,twocolumn,superscriptaddress,amsmath,nofootinbib,10pt,floatfix]{revtex4-1}
\usepackage{graphicx}
\usepackage[normalem]{ulem}
\usepackage{xcolor}
\usepackage{txfonts}
\usepackage{textgreek}
\usepackage[colorlinks, linkcolor=blue, citecolor=blue]{hyperref}
\usepackage{gensymb}
\usepackage[compat=1.1.0]{tikz-feynman}
\tikzset{
pattern size/.store in=\mcSize, 
pattern size = 5pt,
pattern thickness/.store in=\mcThickness, 
pattern thickness = 0.3pt,
pattern radius/.store in=\mcRadius, 
pattern radius = 1pt}

\usepackage{amssymb}
\newcommand{\be}{\begin{equation}}
\newcommand{\ee}{\end{equation}}
\newcommand{\beq}{\begin{eqnarray}}
\newcommand{\eeq}{\end{eqnarray}}
\newcommand{\ba}{\[\begin{aligned}}
\newcommand{\ea}{\end{aligned}\]}
\newcommand{\bal}{\begin{aligned}}
\newcommand{\eal}{\end{aligned}}

\renewcommand{\vec}[1]{{\bf #1}}
\renewcommand{\epsilon}{\varepsilon}

\def\nn{\nonumber}

\newcommand{\avg}[1]{\left\langle #1 \right\rangle}

\renewcommand{\vec}[1]{\boldsymbol{#1}}

\def \k{{\vec{k}}}
\def \q{{\vec{q}}}

\def \r{{\bf {r}}}

\def \tn{\textnormal}

\def \ba{\begin{align*}}
\def \ea{\end{align*}}

\newcounter{indice}

\begin{document}
\title{Intertwined Magnetism and Superconductivity in Isolated Correlated Flat Bands}
\author{Xuepeng Wang} 
\affiliation{Department of Physics, Cornell University, Ithaca, New York 14853, USA.}
\author{J. F. Mendez-Valderrama} 
\affiliation{Department of Physics, Cornell University, Ithaca, New York 14853, USA.}
\author{Johannes S. Hofmann} \email{jhofmann@pks.mpg.de}
\affiliation{Department of Condensed Matter Physics, Weizmann Institute of Science, Rehovot 76100, Israel}
\affiliation{Max-Planck-Institut f\"ur Physik komplexer Systeme,
N\"othnitzer Strasse 38, 01187 Dresden, Germany}
\author{Debanjan Chowdhury}\email{debanjanchowdhury@cornell.edu}
\affiliation{Department of Physics, Cornell University, Ithaca, New York 14853, USA.}
\date{\today}

\begin{abstract}
  Multi-orbital electronic models hosting a non-trivial band-topology in the regime of strong electronic interactions are an ideal playground for exploring a host of complex phenomenology. We consider here a sign-problem-free and time-reversal symmetric model with isolated topological (chern) bands involving both spin and valley degrees of freedom in the presence of a class of repulsive electronic interactions. Using a combination of numerically exact quantum Monte Carlo computations and analytical field-theoretic considerations we analyze the phase-diagram as a function of the flat-band filling, temperature and relative interaction strength. The low-energy physics is described in terms of a set of intertwined orders --- a spin-valley hall (SVH) insulator and a spin-singlet superconductor (SC).  Our low-temperature phase diagram can be understood in terms of an effective SO(4) pseudo-spin non-linear sigma model. Our work paves the way for building more refined and minimal models of realistic materials, including moir\'e systems, to study the universal aspects of competing insulating phases and superconductivity in the presence of non-trivial band-topology. 
\end{abstract}

\maketitle

{\bf Introduction.-} Correlated quantum materials in the intermediate to strong coupling regime often feature a panoply of ordering tendencies leading to complex phase diagrams. The most famous and extensively studied example is that of the cuprate high-temperature superconductor --- a doped Mott insulator which exhibits numerous electronic phases with spontaneously broken symmetries as a result of the frustration between the tendency towards  delocalization and the interaction-induced localization \cite{Keimer15}. While the detailed microscopic mechanisms responsible for the emergence of this complexity is not fully understood, the landscape of competing and intertwined orders has been clarified to a large degree using a variety of points of view \cite{so5_1,SSRMP,LNW,intertwined,PDW}. 

The discovery of two-dimensional moir\'e materials \cite{Andrei2021,Mak2022} has brought a fresh set of challenging theoretical questions to the forefront, involving the physics of interactions projected to a set of isolated nearly flat bands. The projected interactions drive the tendency towards delocalization, as a result of the nontrivial Bloch wavefunctions associated with the flat bands, and localization in the vicinity of commensurate fillings. The quantum geometric tensor associated with these isolated bands is believed to play an important role in much of the essential phenomenology \cite{Bernevigreview,AV20,TS20}. In the absence of a well developed set of theoretical tools and a ``small" parameter that can tackle the generic problem of partially filled, interacting narrow bandwidth (topological) bands, studying even simplified models with carefully designed interactions using complementary techniques can offer new insights and serve as a building block for understanding more realistic models. Specifically, the fate of nearly flat-bands with multiple spin and valley degrees of freedom and projected interactions offers an interesting playground to study the interplay of various ordering tendencies, including superconductivity. 

\begin{figure}[h!]
\includegraphics[width=90mm,scale=1]{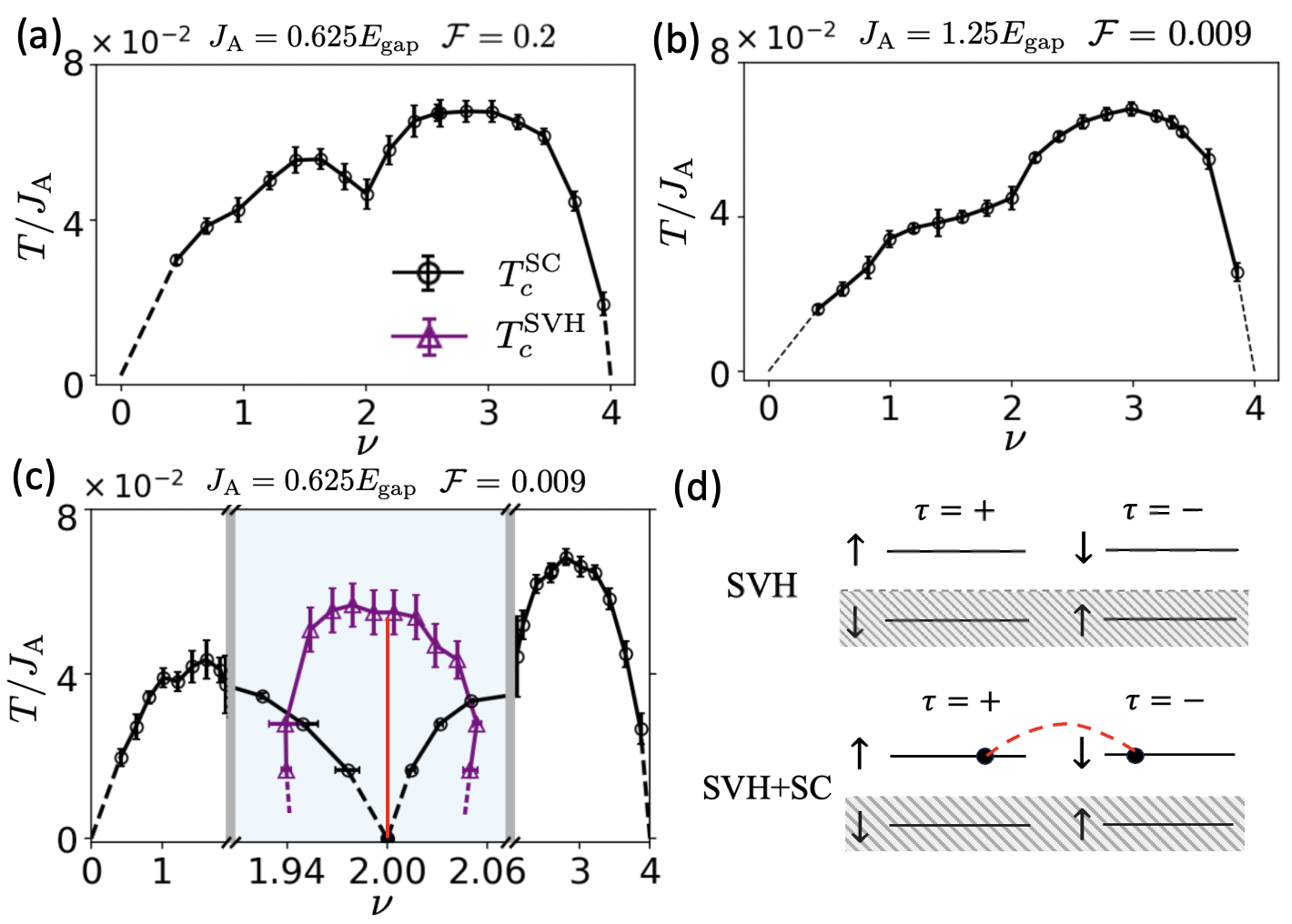}% Here is how to import EPS art
\caption{\label{res_summary} Superconducting phase diagrams for $|J_{\tn{H}}|=J_{\tn{A}}/2$ with: (a) $J_{\tn{A}}/t = 2.5$ and $\mathcal{F}=0.2$, showing a slight suppression of the superconducting $T_c^{\rm{SC}}$ near $\nu=2$, (b)  $J_{\tn{A}}/t = 5$ and $\mathcal{F}=0.009$, and (c) $J_{\tn{A}}/t = 2.5$ and $\mathcal{F}=0.009$, where the zoomed-in blue-shaded region shows the competing SVH insulating state (red solid line) near $\nu=2$. The lightly doped regions near $\nu=2$ exhibit a regime with coexisting SC and SVH order for $T<\min(T_c^{\tn{SC}}, T_c^{\tn{SVH}})$; see Fig.~\ref{0.5flat_result}. (d) The top panel shows a schematic of the SVH insulator at $\nu = 2$ with the shaded regions denoting fully filled electronic bands. The bottom panel is a schematic for the coexisting SVH and SC phases for $\nu = 2+\delta \nu$, where the excess electrons (black circles) form spin-singlet Cooper pairs (red dashed line).}
\end{figure}

With this goal in mind, we will focus on a model of spinful topological (chern) bands that preserve time-reversal symmetry (TRS) and carry a ``valley" degree of freedom. We will study the effect of competing exchange interactions derived, in principle, from a repulsive interaction but designed such that the model does {\it not} suffer from the infamous sign-problem. This will allow us to obtain the phase diagram for the repulsive model over a wide range of temperature, filling, and other microscopic tuning parameters using determinant quantum Monte Carlo (QMC); whereas most of the recent QMC work tied to flat bands has focused on purely attractive interactions \cite{single_chern,peri2021fragile,Zhang2021,chiral,Bernevig21}. Interestingly, we will also be able to obtain the form of the low-energy effective field theory that governs the dynamics and fluctuations tied to the intertwined order-parameters in the projected Hilbert-space, offering complementary analytical insights into the same problem.

{\bf Model. -} We consider a two-dimensional interacting model of topological bands with Chern number, $C=\pm1$, that preserves TRS. The non-interacting bands are obtained microscopically in a model of electrons hopping on the sites of a square lattice \cite{hkin, single_chern}, where the degrees of freedom consist of spin ($\sigma=\uparrow,\downarrow$), valley ($\tau=\pm$) and sublattice ($\eta = \tn{A},\tn{B}$), respectively. The non-interacting part of the Hamiltonian per spin, $H^{(\sigma)}_{\tn{kin}}$, can be written in momentum space as \cite{single_chern},
\beq\label{h_kin0}
H^{(\sigma)}_{\tn{kin}} = \sum_\k\psi^{\dagger}_\k\bigg[B_{0,\k,\sigma}\eta_0 + \vec{B_{\k,\sigma}\cdot\vec{\eta}}\bigg]\tau_0\psi_\k,
\eeq
where $\psi^{\dagger}_\k = (f^{\dagger}_{\k,A,+}, f^{\dagger}_{\k,B,+}, f^{\dagger}_{\k,A,-}, f^{\dagger}_{\k,B,-})$ and $f^{\dagger}_{\k,\eta,\tau}$ denotes the electron creation operator on sublattice $\eta$ with valley $\tau$. Here,  $B_{0,\k,\sigma}$ and $\vec{B}_{\k,\sigma}$ are  matrices determined entirely by the hopping parameters on the underlying lattice, which we assume to include a first ($t$) and a staggered second ($t_2=t/\sqrt{2}$) neighbor hoppings with a $\pi-$flux per square plaquette \cite{si}. Additionally, by including further (e.g. fifth $t_5$) neighbor hoppings, the flatness ratio ${\cal{F}}=W/E_{\tn{gap}}$ ($W\equiv$bandwidth, $E_{\tn{gap}}\equiv$bandgap) can be tuned to be small. By including two copies of $H^{(\sigma)}_{\tn{kin}}$ in a time-invariant fashion \cite{si}, under the operation $\mathcal{T}=i \sigma_y \mathcal{K}$ where $\mathcal{K}$ denotes complex conjugation, we arrive at a model with a set of degenerate topological bands carrying spin and valley with $C=\sigma$. Note that the non-interacting part of the Hamiltonian has a SU(2)$_{\tn{valley}}\times$U(1)$_{\tn{spin}}\times$U(1)$_{\tn{c}}$ symmetry, which is broken {\it explicitly} down to U(1)$_{\tn{valley}}\times$U(1)$_{\tn{spin}}\times$U(1)$_{\tn{c}}$ by the interaction we introduce below.

Our choice of interactions will be inspired by the physics of quantum Hall-type ferromagnetism in spinful Landau levels \cite{gmp,sondhi,qhbilayer,girvin1999quantum}. We will focus on the competing effects of an intra-valley Hund's-type ferromagnetic interaction with $J_{\tn{H}}<0$, and an inter-valley antiferromagnetic interaction with $J_{\tn{A}}>0$, and study the competition between possible valley symmetry-breaking phases and superconductivity in a model with two time-reversed Chern sectors, as introduced above. The interactions take the following form,
\begin{subequations}\label{H_int_tot}
\beq\label{H_int}
H_{\tn{interaction}} &=& H_{\tn{intra-valley}} + H_{\tn{inter-valley}},\\
H_{\tn{intra-valley}} &=& 
J_{\tn{H}}\sum_{\r,\tau=\pm} \vec{S}^{\tau}_{\r}\cdot\vec{S}^{\tau}_{\vec{\r}},\\
%J_{\tn{H}}\sum_{\vec{r},\eta,\tau} \vec{S}^{\eta,\tau}_{\vec{r}}\cdot\vec{S}^{\eta,\tau}_{\vec{r}},\\
H_{\tn{inter-valley}} &=& J_{\tn{A}}\sum_{\vec{r}}\vec{S}^{+}_{\r}\cdot\vec{S}^{-}_{\r}, \label{inter_valley_int}
\eeq
\end{subequations}
where the ``spin" operator is defined as
\beq\label{def_spin}
\vec{S}^{\tau}_{\r} = \sum_{\alpha,\beta = \uparrow,\downarrow}f^{\dagger}_{\r,\tau,\alpha}\vec{\sigma}_{\alpha\beta}f_{\r,\tau,\beta}.
\eeq
We have combined the two-dimensional spatial coordinate and the sublattice index ($\eta$) into $\r$. Note that we have only included an on-site interaction in the full microscopic Hamiltonian (which generates further-neighbor interactions upon projection to the lower flat-bands). 

 {\bf Quantum Monte Carlo.-} The model introduced above possesses an anti-unitary TRS,  $\mathcal{T}'=i \tau_x \sigma_y \mathcal{K}$, which enables a sign-problem-free quantum Monte-Carlo computation at arbitrary filling fraction $\nu$ \cite{congjun,si} as long as $J_{\tn{A}} \geq 2 |J_{\tn{H}}|$. The partition function for the model defined by Eqs.~\ref{h_kin0} and \ref{H_int_tot} is evaluated using a  Trotter decomposition with $\Delta \tau = \beta/N_{\tn{Trotter}}$, where the interaction is factorized via a discrete Hubbard-Stratonovich transformation and the auxiliary fields are sampled stochastically using single spin-flip updates \cite{si}. In the remainder of this manuscript, we will primarily focus on the problem with $J_{\tn{A}} = 2|J_{\tn{H}}|$; this parameter choice corresponds to maximizing the contribution due to Hunds' interaction relative to the antiferromagnetic exchange, where the  Hunds' coupling is expected to be naturally induced from an on-site intra-valley Hubbard repulsion. To address the relative importance of the bare band dispersion vs. (projected) interactions, we will present results for a flatness ratio, $\mathcal{F} = 0.009,~ 0.2$, and $J_{\tn{A}}/t = 2.5 - 5$; the gap to the remote-bands, $E_{\tn{gap}}\simeq 4t$. To obtain the phase diagram as a function of temperature and band-filling, we tune the chemical potential $\mu(T)$ such that $\sum_{\r}\avg{n_{\r}}/L^2 = \nu$, where $\nu$ denotes the filling of the Chern bands and $n_\r=\sum_{\tau,\sigma}f^{\dagger}_{\r,\tau,\sigma}f_{\r,\tau,\sigma}$ is the local electron density ($L^2\equiv$system-size).

{\bf Superconductivity and intertwined orders.-}
Let us begin by discussing the results for the model with $\mathcal{F}=0.2$ ($t_5=0$) and $J_{\tn{A}}/t=2.5$; the bands have some dispersion but the interaction scale is small compared to $E_\tn{gap}$. We find the ground state to be a superconductor over the entire range of fillings, and the transition temperature ($T_c^{\tn{SC}}$) vanishes when $\nu\rightarrow0^+,~4^-$ (see Fig.~\ref{res_summary}a). We compute the temperature-dependent superfluid stiffness, $D_s(T)$, as the transverse electromagnetic response at vanishing Matsubara frequency \cite{criterion},
\beq
D_s = \frac{1}{4}\avg{[-K_{xx}-\Lambda_{xx}(\omega_n=0, q_y=0, q_x \rightarrow 0)]},
\eeq
where $\Lambda_{xx}(\omega_n, \vec{q})$ is paramagnetic current-current correlation, and $K_{xx} \equiv\avg{\partial^2 H[\vec{A}]/\partial A_{x}^2|_{\vec{A}\rightarrow0}}$ is the diamagnetic contribution, with $\vec{A}$ the probe vector potential. The superconducting transition temperature, $T_c^{\tn{SC}}$, is then determined as $T_c^{\tn{SC}} = \pi D_s (T\rightarrow T_c^{\tn{SC}-})/2$ \cite{nelson_kosterlitz}.

Interestingly, we notice a clear suppression of $T_c^{\tn{SC}}$ near $\nu=2$. To compute the tendencies towards pairing and other orders, we introduce the thermodynamic susceptibilities for an observable $O$,
\beq
\chi_O = \frac{1}{L^2}\int d\tau \avg{O^{\dagger}(\tau) O(\tau=0)}.
\eeq
The first observable of interest is associated with a spin-singlet, on-site $s-$wave pairing operator, 
\beq
\Delta_{\tn{SC}}^{\dagger} \equiv \sum_\r [c_{\r,+\uparrow}^{\dagger}c_{\r,-\downarrow}^{\dagger} - c_{\r,+\downarrow}^{\dagger} c^{\dagger}_{\r,-\uparrow}],
\eeq
which also pairs across valleys. The other operator of interest diagnoses the tendency towards an intra-valley ferromagnetic polarization $(\propto [n_{\tau\uparrow} - n_{\tau\downarrow}],~\tau=\pm)$ and an inter-valley antiferromagnetic order $(\propto [n_{+\uparrow} + n_{-\downarrow}])$. We define the associated spin-valley Hall (SVH) order parameter as, 
\beq
\Delta_{\tn{SVH}}\equiv \sum_\r [n_{\r,+\uparrow}+n_{\r,-\downarrow}-n_{\r,+\downarrow}-n_{\r,-\uparrow}].
\eeq
Note that if this observable develops an expectation value near $\nu=2$, it preserves the global TRS. For the data in Fig.~\ref{res_summary}a, we have computed the spin-valley Hall susceptibility, and a finite-size scaling suggests that the system fails to develop this competing order even though $T_c^{\tn{SC}}$ undergoes a downward renormalization (presumably due to enhanced SVH fluctuations) \cite{si}. This is our first indication that SC and SVH orders are intertwined in this model, but depending on values of microscopic parameters one of the two orders becomes energetically favorable.

 To investigate further the possibility of enhancing the tendency to form an insulating ground state at the commensurate filling of $\nu=2$, we focus next on a much flatter band with ${\cal{F}}=0.009$ ($t_5 = (1-\sqrt{2})t/4 $) and two different values of the interaction. For $J_{\tn{A}}/t=5$, the suppression of $T_c^{\tn{SC}}$ near $\nu=2$ disappears, and the ground-state remains a superconductor for all fillings (Fig.~\ref{res_summary}b). In spite of the bands being much flatter, it is worth noting that $J_{\tn{A}}=5t\gtrsim E_{\tn{gap}}$, leading to a mixing with the degrees of freedom from the dispersive remote bands. The model is no longer in the ``projection-only" limit, leading to a reduction in the associated strong-coupling effect tied to just the flat-band Hilbert space\cite{si}. 

\begin{figure}[h!]
\includegraphics[width=90mm,scale=1]{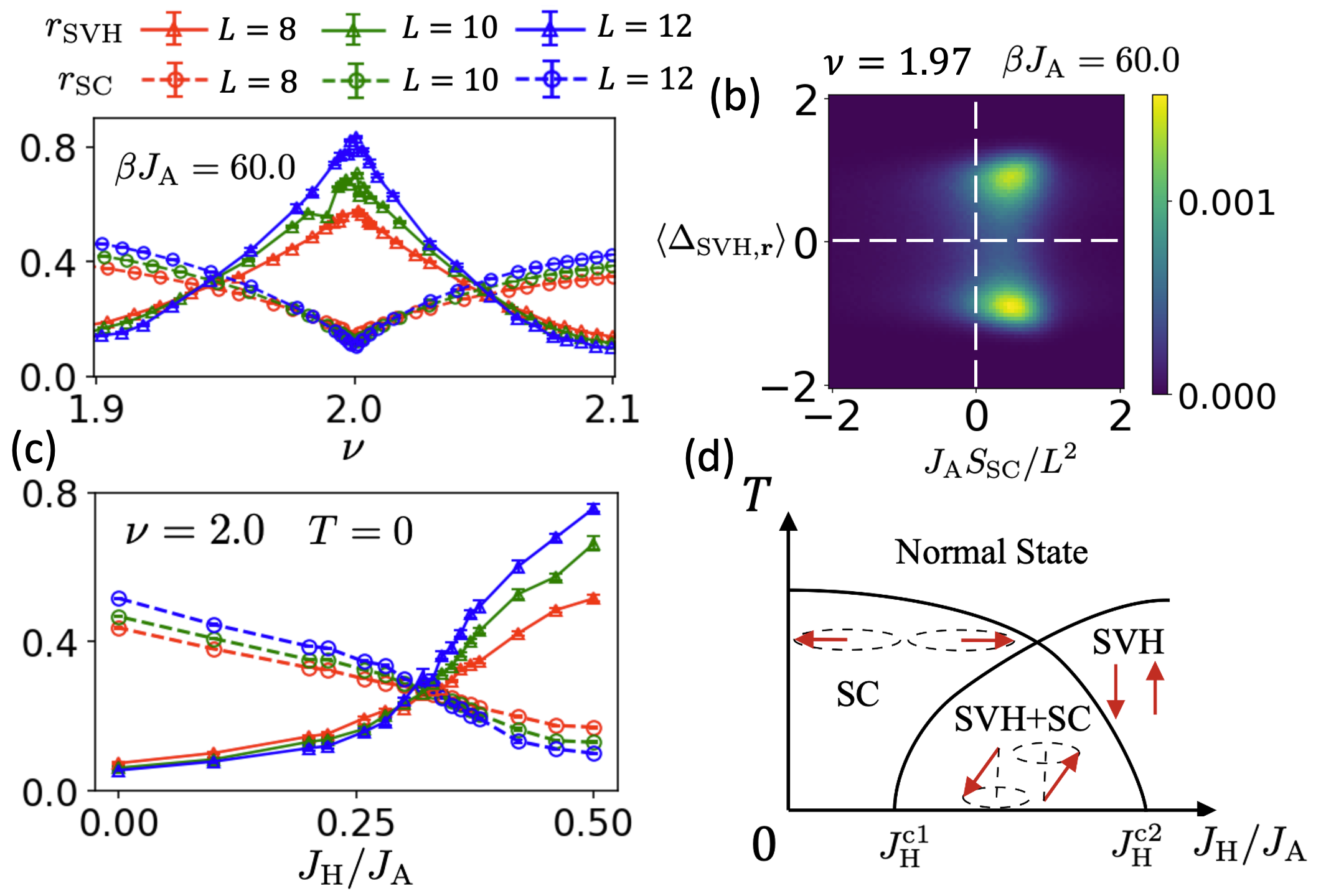}% Here is how to import EPS art
\caption{\label{0.5flat_result} (a) RG-invariant correlation lengths,  $r_{\tn{SVH}}$ (solid line) and $r_{\tn{SC}}$ (dashed line) as a function of $\nu$ at $\beta J_{\tn{A}} =60$ for $J_{\tn{A}}/t = 2.5$ and $\mathcal{F} = 0.009$ with $|J_{\tn{H}}|=J_{\tn{A}}/2$. For both hole  and electron doping relative to $\nu=2$, there exists a regime where both SC and SVH phases coexist for $\nu_{c}^{\tn{SC}}<\nu<\nu_c^{\tn{SVH}}$. (b) Histogram of SVH order parameter, $\avg{\Delta_{\tn{SVH}}}$, and equal-time correlation function, $S_{\tn{SC}}(\q=0)$, measured {\it per Monte-Carlo snapshot} for $\nu=1.97$ and $\beta J_{\tn{A}} = 60$ indicates their microscopic coexistence. (c) The same RG-invariant correlation lengths as a function of $J_{\tn{H}}/J_{\tn{A}}$ at $\nu=2$ and $T=0$. There exists a regime of microscopic coexistence of SC and SVH phases for $J_{\tn{H}}^{\tn{c1}}<J_{\tn{H}}<J_{\tn{H}}^{\tn{c2}}$. (d) A schematic $T-J_{\tn{H}}$ phase-diagram expressed using the pseudo-spin effective model \cite{nelson-fisher1}.}
\end{figure}

Finally, keeping $\mathcal{F}=0.009$ and decreasing the interaction strength to $J_{\tn{A}}=2.5t< E_{\tn{gap}}$, we find a complete suppression of $T_c^{\tn{SC}}\rightarrow0$ at $\nu=2$ (Fig.~\ref{res_summary}c). Using finite-size scaling for $\chi_{\tn{SC}}$ and $\chi_{\tn{SVH}}$, and by carrying out a $T=0$ projective QMC calculation, we provide unambiguous evidence for the ground state being an interaction-induced SVH insulator \cite{si}; see vertical red line in Fig.~\ref{res_summary}c. This indicates that effectively projecting to only the degrees of freedom in the lower ``flatter" bands enhances the commensuration effects at integer filling, in contrast to the previous two cases. We turn next to studying the effect of doping carriers away from the $\nu=2$ insulator on the many-body phase diagram; see Fig.~\ref{res_summary}d.

It is worth noting that while the $\nu=2$ insulator is incompressible, any doping away from this limit will lead to a compressible phase. Moreover, given the prevalence of superconductivity in the model in the absence of the insulating regime, it is likely that the doped model displays superconducting correlations. Thus, the following scenarios for the phase-diagram are possible when $\nu=2\pm \delta\nu$: (i) a first-order transition between the SVH insulator and superconductivity at infinitesimal $\delta\nu$, (ii) phase separation between the SVH insulator and SC over a range of intermediate $\delta\nu$, and (iii) a phase with microscopically co-existent SVH order and SC. To diagnose the competition between SVH and SC phases, we focus on the renormalization group (RG)-invariant correlation length, $r_O$, obtained from the equal-time correlation function $S_O(\q)$ as
\begin{subequations}
\beq
&
r_O \equiv \frac{\xi_O}{L} = \frac{1}{2L \sin(\pi/L)}\sqrt{\frac{S_O(\vec{q} = \vec{0})}{S_O(\vec{q} = (2\pi/L,0))} - 1},\\&
S_O(\q) = \frac{1}{L^2}\sum_{\r,\r'}e^{-i(\r-\r')\cdot\q}\avg{O^{\dagger}(\r')O(\r)}.
\eeq
\end{subequations}

By extracting $T_{c}^{\tn{SVH}}$ (purple triangles) and $T_c^{\tn{SC}}$ (black circles) in Fig.\ref{res_summary}c at a fixed filling in the vicinity of $\nu=2$, we find that both orders are present below $T<\min(T_{c}^{\tn{SVH}},T_{c}^{\tn{SC}})$. We note that the slight ``bending'' of $T_{c}^{\tn{SVH}}$ with decreasing temperature inside the superconducting phase is likely due to the competition between the two orders.

We have analyzed the correlation length for SVH and SC as a function of filling fraction $\nu$ at a fixed temperature $\beta J_{\tn{A}} = 60.0$, as shown in Fig.\ref{0.5flat_result}a.  Our finite size scaling analysis suggests that there exists a range of $\delta\nu$ on either side of $\nu=2$, where SVH order survives and $T_c^{\tn{SC}}\propto|\delta\nu|$. 

To address the question of microscopic co-existence vs. phase separation, we analyze the histogram of SVH order parameter, $\avg{\Delta_{\tn{SVH}}}$, and equal-time correlation function, $S_{\tn{SC}}(\q=0)$, measured {\it per Monte-Carlo snapshot}, instead of ensemble-averaged observables. The histogram for $\nu=1.97$ and $\beta J_{\tn{A}} = 60$, is shown in Fig.~\ref{0.5flat_result}b. If the system had a tendency to phase-separate, the Monte-Carlo snapshots would show \textit{either} SVH \textit{or} SC order, appearing as ``blobs" along the axes in the histogram.  On the other hand, an off-diagonal peak of the histogram in Fig.\ref{0.5flat_result}b suggests that within each Monte-Carlo snapshot, both SVH and SC orders co-exist, indicating a coexistence between SVH and SC orders for $\nu=1.97$ and $\beta J_{\tn{A}} = 60$.

Instead of tuning the filling near $\nu=2$ at a fixed value of $|J_{\rm{H}}|/J_{\rm{A}}$ and driving transitions between the different phases, we can gain complementary insights into the strong-coupling limit by varying the ratio of interactions at a fixed $\nu=2$. A finite-size scaling analysis \cite{fakher_scaling} of the correlation length ratios obtained from our $T=0$ projective simulations are shown in Fig.\ref{0.5flat_result}c. At $\nu=2$ we find a coexistence of SVH and SC orders  between $|J_{\tn{H}}^{\tn{c1}}|/J_{\tn{A}}=0.295(3)$ and $|J_{\tn{H}}^{\tn{c2}}|/J_{\tn{A}}=0.342(4)$. To help unify our understanding of these competing phases and their phase-transitions, let us turn next to an analytical approach that helps tie together the numerical phenomenology.

{\bf Analytical Results.-} Given the orders we found in our QMC computations, it is natural to address the nature of the effective field theory for a ``super-spin" \cite{so5_3,ashvin_skrymion,SS20} that describes the phases and possible phase-transitions in the low-energy Hilbert space; such approaches have been used earlier in the context of orders in the cuprates \cite{CHN,so5_2,so6,Efetov2013}. We introduce the Nambu spinors, $\Psi^{\dagger} \equiv (c^{\dagger}_{+\uparrow}, c^{\dagger}_{-\uparrow}, c_{+\downarrow}, c_{- \downarrow})$, and the Pauli matrices, $\mu_\alpha$, which act on the particle-hole subspace. The three-dimensional pseudo-spin vector operator, $\vec{n} \equiv (\tn{Re}[\Delta_{\tn{SC}}], -\tn{Im}[\Delta_{\tn{SC}}],\Delta_{\tn{SVH}})$, encodes the competing orders of interest and can be expressed as bilinears of $\Psi$:
\beq\label{na_main}
&&n_1 = \frac{1}{2}\Psi^{\dagger} \mu_x \tau_x \Psi\equiv \tn{Re}[\Delta_{\tn{SC}}],\nn\\
&&n_2 =\frac{1}{2}\Psi^{\dagger} \mu_y \tau_x \Psi \equiv -\tn{Im}[\Delta_{\tn{SC}}],\nn\\
&& n_3 = \frac{1}{2}\Psi^{\dagger} \tau_z \Psi \equiv \Delta_{\tn{SVH}}. 
\eeq
Naively, one might expect the low-energy theory to be described purely in terms of $\vec{n}$, until one notices that these set of operators do not form a closed group. The minimal group that contains the \{$n_\alpha$, $\alpha=1,2,3$\} as generators, is SO(4). The remaining three generators, $\{L_\alpha \}$ act as the angular momentum of $n_\alpha$ and are given by, 
\beq\label{La_main}
%\begin{aligned}
&&L_1 = -\frac{1}{2}\Psi^{\dagger} \mu_y \tau_y \Psi\equiv  \tn{Re}[\Delta_{\tn{vSC}}],\nn\\ 
&&L_2 = \frac{1}{2}\Psi^{\dagger} \mu_x \tau_y \Psi\equiv -\tn{Im}[\Delta_{\tn{vSC}}],\nn\\
&& L_3 = \frac{1}{2}\Psi^{\dagger} \mu_z \Psi =  \frac{1}{2}\bigg(\sum_{\tau, \sigma} n_{\tau, \sigma} - 2\bigg),   
%\end{aligned}
\eeq
where the additional order parameter,
\beq
\Delta_{\r,\tn{vSC}}\equiv [c_{\r,-\downarrow}c_{\r,+\uparrow} + c_{\r,-\uparrow}c_{\r,+\downarrow}].
\eeq
The above pseudo-spin operators can be mapped to the spin operators on a bipartite lattice \cite{nelson-fisher1,nelson-fisher2} where the $L_\alpha$ represent the uniform magnetization, while the $n_\alpha$ are equivalent to the staggered magnetization in the $\alpha$ direction of the spin model, respectively \cite{si}.

Projecting the interactions in Eq.~\ref{H_int_tot} to the lower flat-bands, we obtain the following low-energy effective Hamiltonian,
\begin{equation}
\begin{aligned}
&{\cal{H}}_{\tn{eff}} =\sum_{\r,\alpha} \bigg[G \bigg( U_\alpha L_{\alpha,\r}^2 + V_\alpha n_{\alpha, \r}^2 \bigg) + \mu L_{3, \r}\bigg] \\& 
~~~~~~~~~~+\sum_{\r,\r',\alpha}F_{\r,\r'}\bigg[U'_{\alpha}L_{\alpha,\r}L_{\alpha,\r'} +V'_{\alpha} n_{\alpha,\r}n_{\alpha,\r'}\bigg]\label{proj_ham_nlr},
\end{aligned}
\end{equation}
where the coefficients are given by,
\begin{subequations}
\beq
&
U_1 =U_2 = \frac{J_A}{5}+\frac{3|J_H|}{10}, ~~U_3= -\frac{J_A}{20}+\frac{3|J_H|}{10}\\&
V_1=V_2=-\frac{3J_A}{10}+\frac{3|J_H|}{10}, ~~V_3 = -\frac{3J_A}{10}-\frac{6|J_H|}{5}\\&
U'_1 =U'_2 = \frac{J_A}{8}, ~~U'_3= \frac{3|J_H|}{4}\\&
V'_1=V'_2=-\frac{3J_A}{8}, ~~V'_3 = -\frac{J_A}{4} - \frac{3|J_H|}{4}.
\eeq
\end{subequations}
Here, $\mu$ acts as a pseudo magnetic-field. Note that the model does not have any (emergent) SO(4) symmetry, but it nevertheless provides an organizing framework to describe the various order-parameter fluctuations. The coefficients $G$ and $F_{\r,\r'}$ are positive and can be obtained in terms of the Wannier functions constructed out of the lower flat-band Bloch wavefunctions; their precise form is unimportant for describing the phases and phase-transitions at $T=0$, which we turn to next.

When $\mu=0$, the effective Hamiltonian in Eq.\ref{proj_ham_nlr} hosts an anisotropy-tuned easy-axis to easy-plane transition with decreasing $|J_{\tn{H}}|/J_{\tn{A}}$, as seen in our numerical data in Fig.~\ref{0.5flat_result}c. The pseudospin-flop transition has been studied theoretically in classic papers \cite{nelson-fisher1,nelson-fisher2}; a schematic phase diagram as a function of $J_{\tn{H}}/J_{\tn{A}}$ and $T$ appears in Fig.~\ref{0.5flat_result}d. Let us now elaborate further on the connections between the pseudospin model and the numerically obtained phase-diagram at $T=0$. For $\nu=2$, the uniform polarization $\avg{L_3}= 0$ vanishes across the entire phase diagram. When $J_{\tn{H}}/J_{\tn{A}}=0$, the competition between the isotropic on-site term and the long-range interactions generated by $F_{\r,\r'}$ lead to an easy-plane Neel state with $\avg{n_{1,2}}\neq0$, suggesting that the ground state is a spin-singlet superconductor (Fig.~\ref{0.5flat_result}d).

\begin{figure}[h!]
\includegraphics[width=80mm,scale=1]{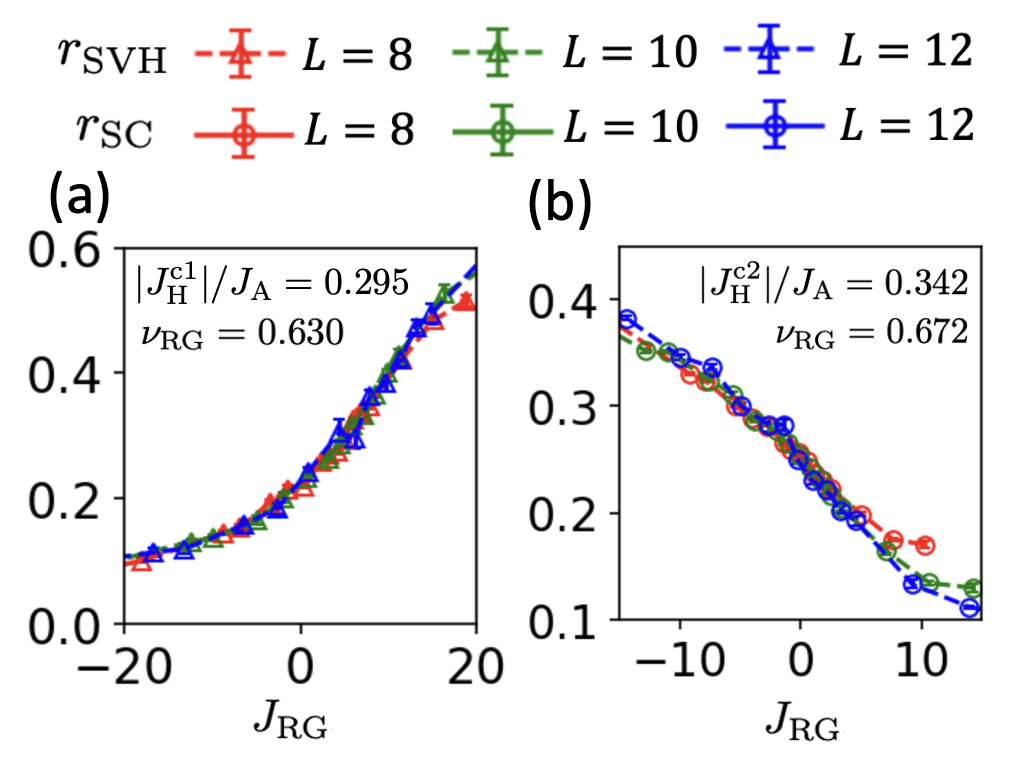}% Here is how to import EPS art
\caption{\label{scaling} Scaling collapse analysis of the anisotropy-tuned transition at $T=0$ and $\nu=2.0$ for (a) the onset of SVH order, described by a $(2+1)-$dimensional Ising theory, (b) the loss of SC order, associated with a $(2+1)-$dimensional XY transition. The RG invariant tuning parameter is defined as $J_{\tn{RG}}=[(J_{\tn{H}}-J_{\tn{H}}^{\tn{c}})/J_{\tn{H}}^{\tn{c}}]L^{1/\nu_{\tn{RG}}}$, with $J_{\tn{H}}^{\tn{c}}$ labeled in the respective panels.}
\end{figure}

The gapped SC ground state remains stable with increasing $|J_{\tn{H}}| < |J_{\tn{H}}^{\tn{c1}}|$, across which there appears a continuous quantum phase transition to a phase with coexisting SVH and SC orders. In the pseudospin language, for $|J_{\tn{H}}^{\tn{c1}}|< |J_{\tn{H}}|<|J_{\tn{H}}^{\tn{c2}}|$, they tilt away from the $xy-$plane, such that both the SC order parameter $\avg{n_{1,2}}\neq0$ and SVH order parameter $\avg{n_{3}}\neq0$. Increasing $|J_{\tn{H}}|$ beyond $|J_{\tn{H}}^{\tn{c2}}|$ turns the easy-plane anisotropy in Eq.\ref{proj_ham_nlr} to an easy-axis anisotropy, where SC disappears ($\avg{n_{1,2}}=0$) and only the SVH order parameter survives  $\avg{n_{3}}\neq0$. The chemical potential tuned transitions between the SVH and SC phases can also be described within the above picture in terms of an external magnetic-field tuned pseudospin-flop transition \cite{nelson-fisher1,nelson-fisher2}; see \cite{si} for a detailed discussion of the differences from the present case.

To finally address the universality class associated with the distinct anisotropy-tuned phase transitions \cite{nelson-fisher1,nelson-fisher2} for $T=0$ and $\nu=2$ at $J_{\tn{H}}^{\tn{c1}}$ and $J_{\tn{H}}^{\tn{c2}}$, we perform a scaling collapse analysis in Fig.~\ref{scaling}. The onset of SVH order in the presence of a background SC order at $J_{\rm{H}} = J_{\rm{H}}^{c1}$ (where the fermions are already gapped) belongs in the $(2+1)$-dimensional Ising universality class; the correlation-length critical exponent in Fig.~\ref{scaling}a is consistent with Ising criticality. Similarly, the loss of SC at $J_{\rm{H}} = J_{\rm{H}}^{c2}$ in the absence of any gapless fermions belongs in the $(2+1)$-dimensional XY universality class, as can be seen from the rescaled data in Fig.~\ref{scaling}b.

{\bf Outlook.-} We have studied the effects of competing intra-valley ferromagnetic and inter-valley antiferromagnetic interactions --- derived from a purely repulsive electronic interaction --- projected to a set of isolated topological flat bands. The low-energy physics is described by a set of intertwined orders involving a spin-valley Hall insulator and superconductor near the commensurate filling of $\nu=2$.
Clearly, this competition also rules out the prospect of any applicable {\it lower} bounds on the superconducting $T_c^{\rm{SC}}$ \cite{chiral}, in contrast with analogous suggestions for models with on-site attractive interactions \cite{dsbound3,Bernevig21, dsbound1}. 
In the symmetry broken phases, we have also identified the effective field theory for the intertwined orders, and pinned down the universal theories for the associated phase transitions. The universal physics near the finite temperature multicritical point deserves a more careful study in the future, as the normal metallic phase without any symmetry-breaking orders involves gapless fermions coupled to the critical order-parameter fluctuations.  

Our findings have a number of conceptual similarities with the phenomenology of correlated insulators and superconductivity, when doped away from the commensurate fillings, in moir\'e graphene.  A recent moir\'e-inspired numerical study has also highlighted the role of competing orders for models of topological multi-orbital flat-bands with the full repulsive density-density interactions \cite{zaletel_dmrg}. It has not escaped our attention, that our model shares superficial similarities with a model displaying skyrmion mediated pairing as well \cite{so5_3,ashvin_skrymion,SS20}. However, our current model does not yield skyrmionic excitations within a Chern sector as the cheapest excitation. Other variations of the above model may be able to host a quantum Hall-like ferromagnetic ground state at integer filling and possibly a skyrmion-mediated superconducting phase, which we leave for future study.
In addition,  investigating various proxies for electrical transport near the symmetry-breaking transitions remains an exciting avenue.

{\bf Acknowledgements.-} We thank Erez Berg for a number of interesting discussions and related collaborations. JH was supported by the European Research Council (ERC) under grant HQMAT (grant no. 817799), the US-Israel Binational Science Foundation (BSF), and a Research grant from Irving and Cherna Moskowitz. This work is supported in part by a Sloan research fellowship from the Alfred P. Sloan foundation to DC.
This work used Expanse at the San Diego Supercomputer Center through allocation TG-PHY210006 from the Advanced Cyberinfrastructure Coordination Ecosystem: Services \& Support (ACCESS) program \cite{access}, which is supported by National Science Foundation grants \#2138259, \#2138286, \#2138307, \#2137603, and \#2138296.
The auxiliary field QMC simulations were carried out using the ALF package \cite{alf}.
\bibliographystyle{apsrev4-1_custom}
\bibliography{draftv1.bib}

\clearpage
\input{supplementary_core}

\end{document}

%% file: supplementary_core.tex
\renewcommand{\thefigure}{S\arabic{figure}}
\renewcommand{\figurename}{Supplemental Figure}
\setcounter{figure}{0}
%\setcounter{page}{0}
%\appendix
\begin{widetext}
{\bf SUPPLEMENTARY INFORMATION for ``Intertwined Magnetism and Superconductivity in Isolated Correlated Flat Bands"}\\
\begin{center}
    X. Wang,~J.F. Mendez-Valderrama,~J.S. Hofmann,~D. Chowdhury
\end{center}

\section{Band dispersion for topological bands}
As introduced in the main text, the matrices in Eq. \ref{h_kin0} that determine the dispersions for the bands are given by,

\begin{equation}
\begin{aligned}
B^{x}_{\k,\uparrow} + iB^{y}_{\k,\uparrow} &= -2t \bigg[e^{-i\frac{\pi}{4}-ik_y}\cos(k_y) + e^{i\frac{\pi}{4}-ik_y}\cos(k_x)\bigg] \\
B^{z}_{\k,\uparrow} &= -2t_2 \bigg[ \cos(k_x + k_y) - \cos(k_x - k_y) \bigg] \\
B^{0}_{\k,\uparrow} &= -2t_5 \bigg[ \cos(2k_x + 2k_y) + \cos(2k_x - 2k_y) \bigg]
\end{aligned},
\end{equation}
where $t,~t_2,~t_5$ are the hopping parameters. For the simulations with $\mathcal{F}=0.009$, the hopping parameters are given by $t_2 = t/\sqrt{2}$ and $t_5 = (1-\sqrt{2})t/4 $ \cite{single_chern}; whereas for the simulations with $\mathcal{F}=0.2$, the hopping parameters are given by  $t_2 = t/\sqrt{2}$ and $t_5 = 0 $, respectively. The non-interacting band structure and associated Berry curvature distribution in the Brillouin zone are shown in Fig.~\ref{non_interacting_band}.
\begin{figure}[h!]
\includegraphics[width=180mm,scale=1]{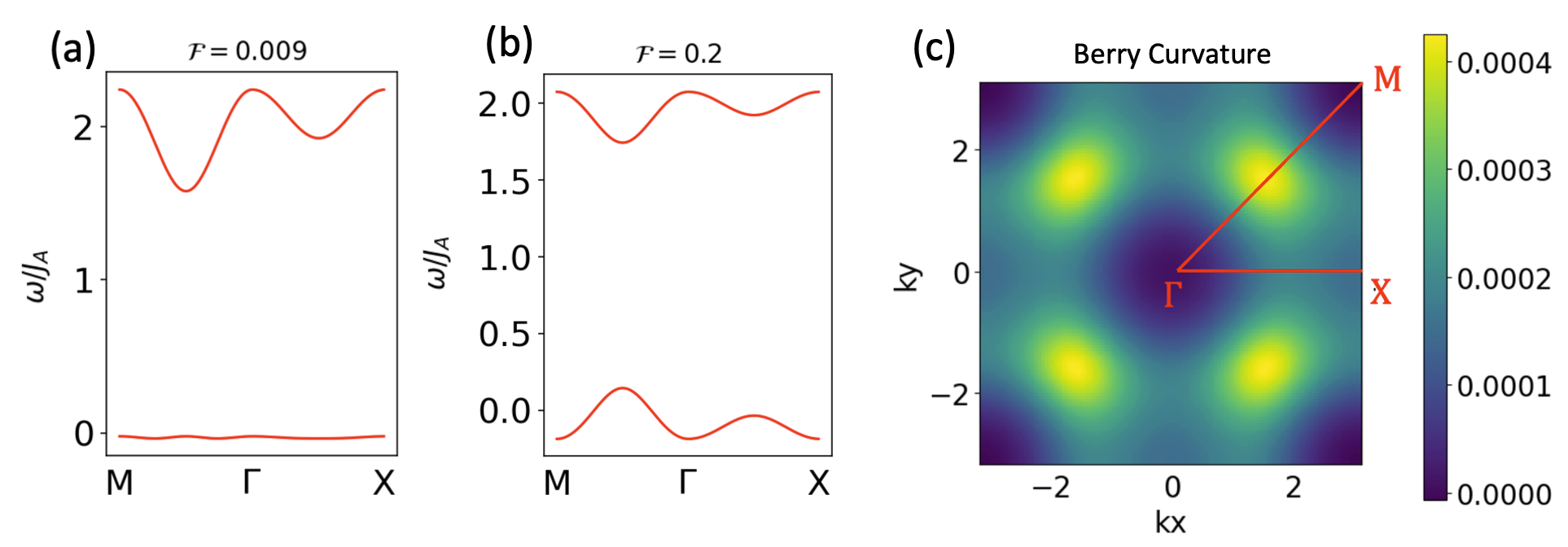}
\caption{\label{non_interacting_band} Non-interacting band structure for (a) $\mathcal{F} = 0.009$, and (b) $\mathcal{F} = 0.2$, respectively. (c) Berry curvature distribution, which is identical for $\mathcal{F} = 0.009$ and $\mathcal{F} = 0.2$.}
\end{figure}

\section{Sign-Problem-Free simulation}\label{App::sign problem}
In this section, we discuss the origin of the sign-problem-free nature of the model defined in Eqn.~\ref{h_kin0} and \ref{H_int_tot}, respectively. For the auxiliary field quantum Monte-carlo simulations, the presence of an anti-unitary symmetry, $\mathcal{T}' = i\tau_x \sigma_y\mathcal{K}$, is key \cite{congjun}. Let us re-write the interaction  Hamiltonian, $H_{\tn{interaction}}$, in Eqn.~\ref{H_int_tot} as,
\beq
H_{\tn{interaction}} &= \sum_{\vec{r}} \bigg[J_1 (\vec{S}^{+}_{\vec{r}} + \vec{S}^{-}_{\vec{r}})^2 - J_2 (\vec{S}^{+}_{\vec{r}} - \vec{S}^{-}_{\vec{r}})^2 \bigg],
\eeq
where $J_1 = J_{\tn{A}}/4 + J_{\tn{H}}/2$ and $J_2 = J_{\tn{A}}/4 - J_{\tn{H}}/2$, respectively. Introducing a Hubbard-Stratonovich decomposition using auxiliary fields, $\vec{\phi}_{1,\r}$, $\vec{\phi}_{2,\r}$, in the spin channel, at leading order in $\Delta \tau$, the imaginary-time evolution operator is given by,
\beq
-\Delta\tau H_{\tn{interaction}} &= \sum_{\vec{r}} \bigg[\sqrt{-\Delta\tau J_1} ~ \vec{\phi}_{1,\r} \cdot (\vec{S}^{+}_{\vec{r}} + \vec{S}^{-}_{\vec{r}}) - i \sqrt{-\Delta\tau J_2} \vec{\phi}_{2,\r} \cdot(\vec{S}^{+}_{\vec{r}} - \vec{S}^{-}_{\vec{r}})\bigg].
\eeq
The spin operators in each valley transform under $\mathcal{T}'$ as $\mathcal{T}' \vec{S}^{\pm}_{\vec{r}} \mathcal{T}'^{-1} = -\vec{S}^{\mp}_{\vec{r}}$. It is straightforward to show that if $J_1>0$ and $J_2>0$, $-\Delta\tau H_{\tn{interaction}}$ is invariant under $\mathcal{T}'$, which ensures the eigenstates of $H_{\tn{interaction}}$ can be grouped into Kramer doublets ensuring the positive-definiteness of the partition function \cite{congjun}. Similar argument also holds for $H_{\tn{kin}}$, and thus the model is sign-problem-free.

\section{Superfluid stiffness, critical temperature ($T_c^{\tn{SC}}$), and critical filling ($\nu_c^{\tn{SC}}$)}\label{App::Ds}

In this section, we provide additional details for determining the superconducting transition temperature $T_c^{\tn{SC}}$ at a fixed filling fraction $\nu$ and the critical filling fraction $\nu_c^{\tn{SC}}$ at the two lowest temperatures in our QMC calculations. 

We determine $T_c^{\tn{SC}}$ (denoted by purple stars) from the temperature dependent superfluid stiffness, $D_s(T)$, as shown in Fig.~\ref{Tc_sc_dome}. To extract the critical temperature \cite{nelson_kosterlitz}, we fit a smooth quadratic polynomial (blue solid line) to the superfluid stiffness $D_s(T)$ near the crossing with the line, $2T/\pi$ (black dashed line). The resulting errors in determining $T_c^{\tn{SC}}$ are almost negligible. 

\begin{figure}[h!]
\includegraphics[width=180mm,scale=1]{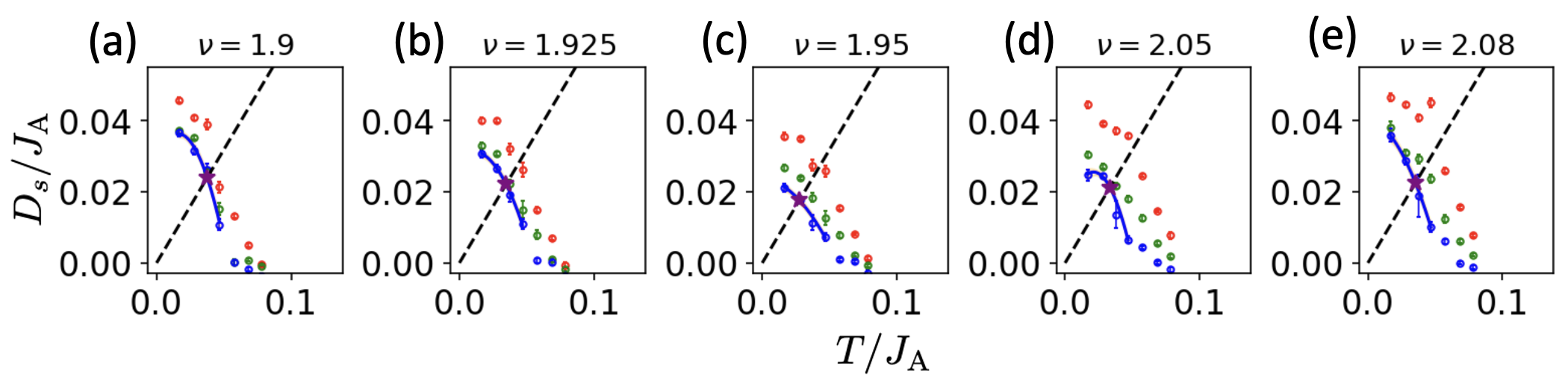}
\caption{\label{Tc_sc_dome} Temperature dependence of the superfluid stiffness, $D_s(T)$, at fixed filling fractions shown in panels (a)-(e). Black dashed line represents $2T/\pi$. The colored data points represent the different system sizes, as in the main text.}
\end{figure}

Similarly, $\nu_c^{\tn{SC}}$ is determined from the criterion $D_s(\nu_c^{\tn{SC}}, T) = 2 T/\pi $ at a fixed temperature $T$. The filling fraction dependence of $D_s(\nu)$ at fixed temperature for $\beta J_\tn{A} = 35.8$ and $\beta J_\tn{A} = 60.0$ are shown in Fig.~\ref{nc_sc}a and Fig.~\ref{nc_sc}d, respectively. A clear suppression of $D_s(\nu)$ can be observed at $\nu = 2$. We use a similar quadratic polynomial fit for the stiffness for $L=12$. The $\nu_c^{\tn{SC}}$ are determined from the crossing points (red data-points with error-bar); the yellow shaded regions in Fig.~\ref{nc_sc} represent the error generated in the fitting process.

\begin{figure}[h!]
\includegraphics[width=150mm,scale=1]{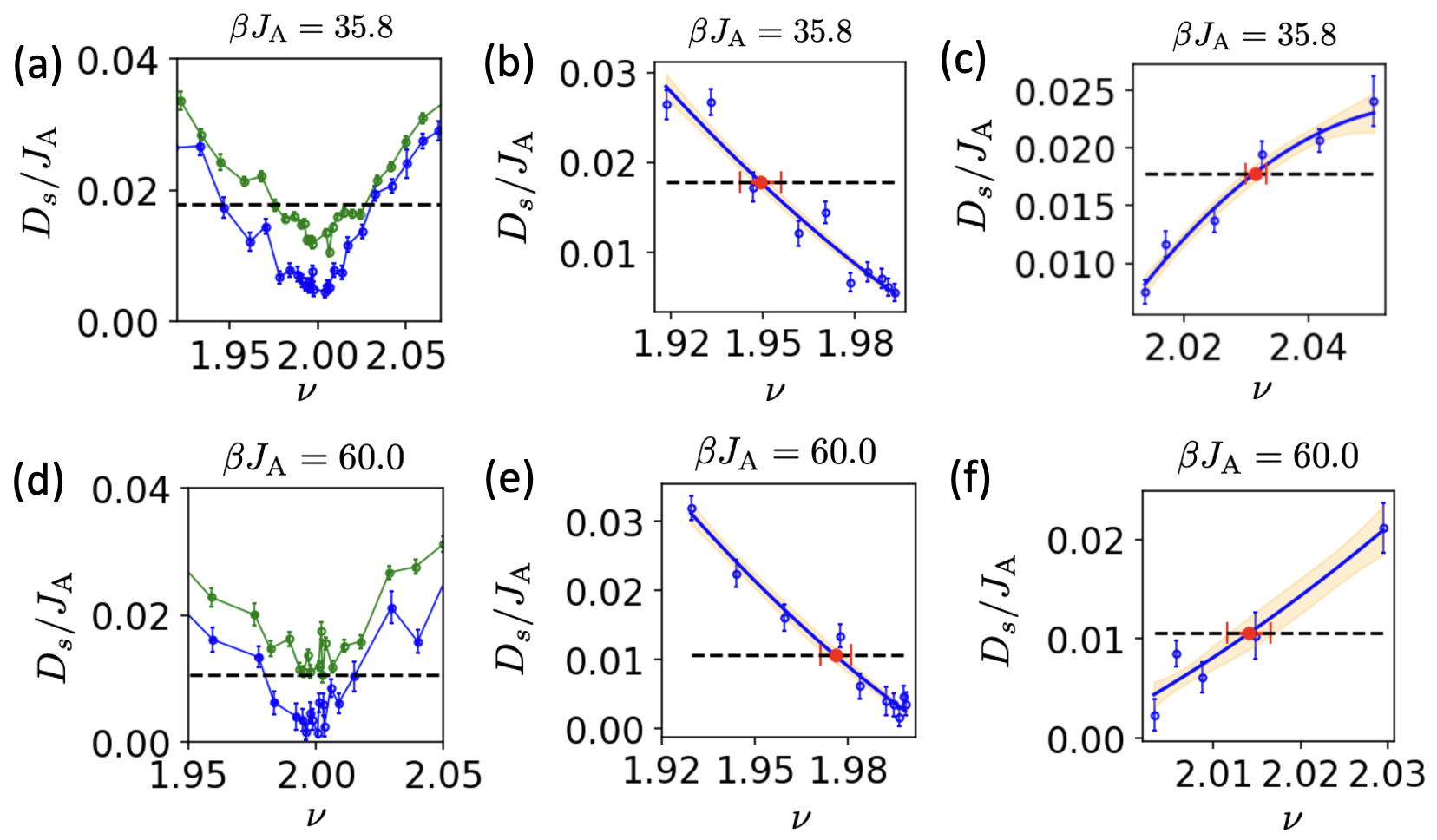}
\caption{\label{nc_sc} (a) and (d) Superfluid stiffness, $D_s$, as a function of $\nu$ at two different temperatures. The black dashed line represents $2T/\pi$ and colors denote different system sizes (see main text). Least square fit of $D_s(\nu)$ as a function of $\nu$, with fitting error denoted by the yellow shaded region. Red data-point represents the critical filling fraction $\nu_c^{\tn{SC}}$ for (b-c) $\beta J_\tn{A} = 35.8$, and (e-f) $\beta J_\tn{A} = 60.0$. }
\end{figure}

Finally, we have also performed a finite-size-scaling analysis of the RG-invariant correlation length, $r_{\tn{SC}}\equiv \xi_{\tn{SC}}/L$, to obtain $\nu_c^{\tn{SC}}$ independently (i.e. instead of using the BKT criterion) from the crossing point (black star in Fig.~\ref{nc_sc2}) between the two largest system sizes. 

\begin{figure}[h!]
\includegraphics[width=170mm,scale=1]{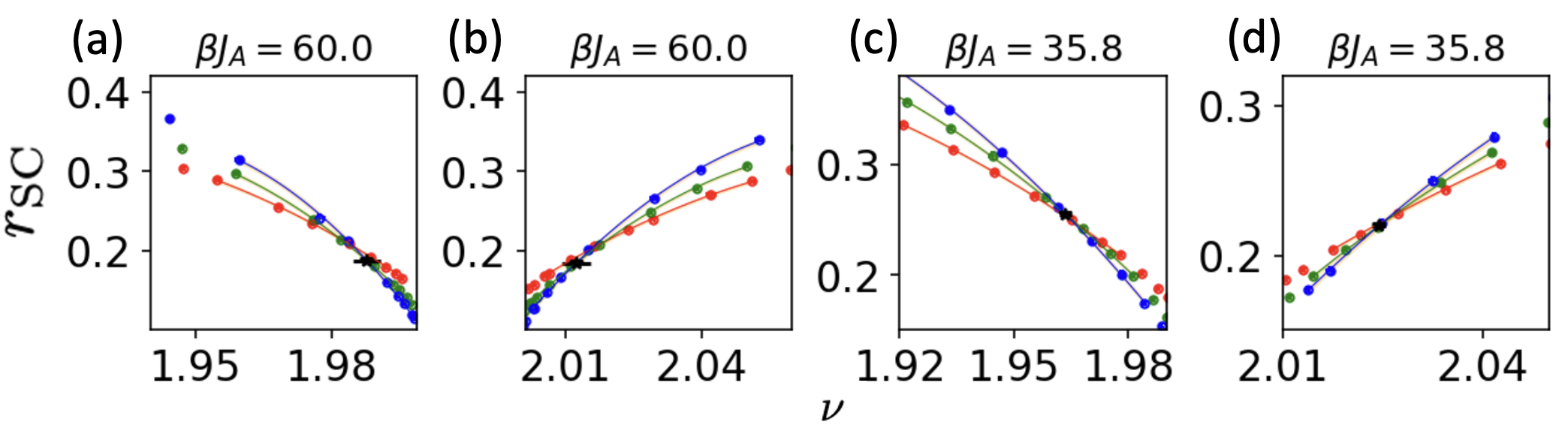}% Here is how to import EPS art
\caption{\label{nc_sc2} RG-invariant correlation length, $r_{\tn{SC}}$, as a function of $\nu$ at two different temperatures. Solid lines denote the least squares fit, with the yellow shaded region showing the associated errors. The critical $\nu_c^{\tn{SC}}$ is determined from the crossing point of the two largest system sizes.}
\end{figure}

\section{Effect of remote bands}\label{App::remote}

In this section, we provide more details comparing the results in Fig.\ref{res_summary} in the aspect of the validity of the `projection-only' limit. We define the electron occupation projected to the remote bands $\avg{n_{\tn{remote}}(\k)}$ as
\beq
\avg{n_{\tn{remote}}(\k)} \equiv \sum_{\eta\tilde{\eta},\alpha\beta,\tau}G^{\tn{eq}}_{\eta\alpha;\tilde{\eta}\beta,\tau}(\k) u_{\eta\alpha}(\k) ~u^{*}_{\tilde{\eta}\beta}(\k)
\eeq
, where $G^{\tn{eq}}_{\eta\alpha;\tilde{\eta}\beta,\tau}(\k) \equiv \avg{c^\dagger_{\eta\alpha,\tau}(\k) c_{\tilde{\eta}\beta,\tau}(\k)}$ denotes the equal-time Green's function obtained from QMC and $u_{\eta\alpha}(\k)$ denotes the wavefunction of the remote band with $\eta\tilde{\eta}$ as sublattice index, $\alpha \beta$ as spin index and $\tau$ as valley index. In Fig.\ref{n_remote}, we plot $\avg{n_{\tn{remote}}(\k)}$ along the high-symmetry lines in BZ for three cases we studied in the main text. For $J_{\tn{A}}/E_{\tn{gap}} = 0.625$, $\avg{n_{\tn{remote}}(\k)}$ is negligible regardless of the flatness ratio $\mathcal{F}$; in comparison, for $J_{\tn{A}}/E_{\tn{gap}} = 1.25$ (blue curve), $\avg{n_{\tn{remote}}(\k)}$ is significantly enhanced, indicating the interaction starts to `mix' between the active bands and remote bands.

\begin{figure}[h!]
\includegraphics[width=160mm,scale=1]{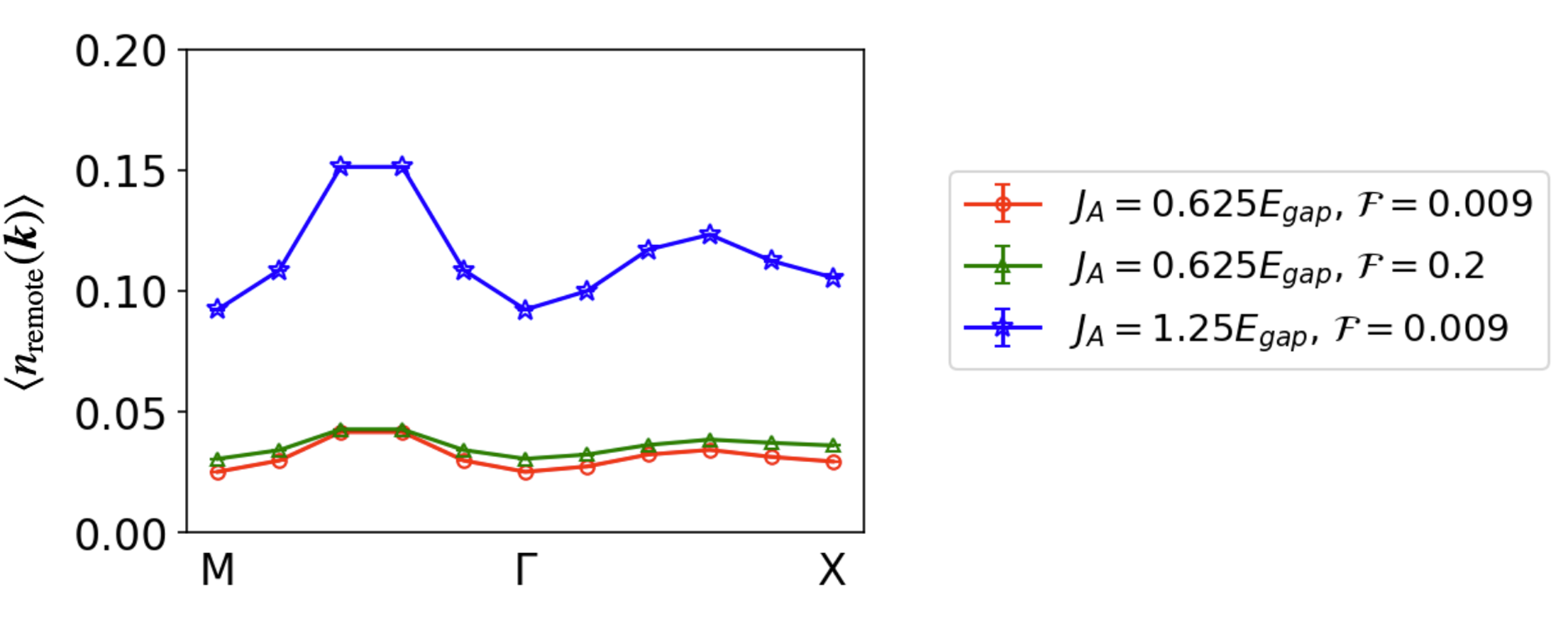}% Here is how to import EPS art
\caption{\label{n_remote} Electronic occupation projected to the remote bands $\avg{n_{\tn{remote}}(\k)}$, at $\beta J_{\tn{A}} = 60.0$ and $\nu=2.0(1)$ with a system size $L=10$ (error-bars are smaller than the symbols). A large $J_{\tn{A}}/E_{\tn{gap}}$ will increase the remote band occupation regardless of the flatness ratio.}
\end{figure}

\newpage
\section{Determination of $T_{c}^{\tn{SVH}}$ and $
\nu_{c}^{\tn{SVH}}$}\label{App::proj}

We now provide additional details for determining the SVH order critical filling, $\nu_c^{\tn{SVH}}$ (Fig.\ref{nc_svh}), and critical temperature, $T_c^{\tn{SVH}}$ (Fig.\ref{Tc_svh}). Both are obtained from a finite-size scaling analysis of the RG-invariant correlation length, $r_\tn{SVH}\equiv \xi_{\tn{SVH}}/L$. 

The temperature dependence of $r_\tn{SVH}(T)$ at fixed filling fraction, $\nu$, are shown in Fig.\ref{Tc_svh}. $T_c^{\tn{SVH}}$ is determined from the crossing point (denoted by black star) for different system sizes, and the error is determined by the temperature difference between the closest data-points near $T_c^{\tn{SVH}}$. To pinpoint $\nu_c^{\tn{SVH}}$, a least-square fit of $r_{\tn{SVH}}(\nu)$ is applied as a function of $\nu$ for each system size (Fig.~\ref{nc_svh}), and the crossing point (black star) for the largest two system sizes is determined.

\begin{figure}[h!]
\includegraphics[width=160mm,scale=1]{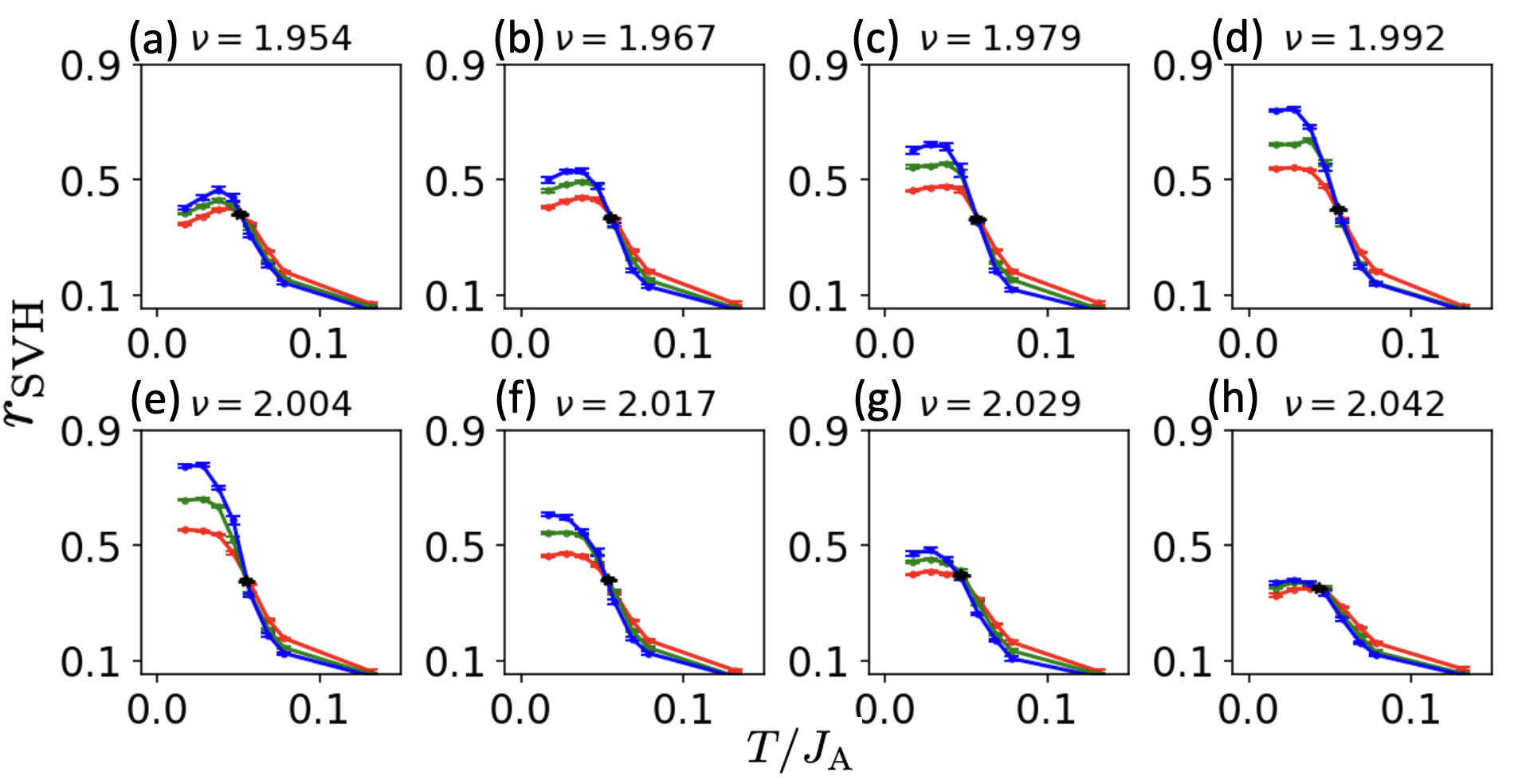}% Here is how to import EPS art
\caption{\label{Tc_svh} RG-invariant correlation length, $r_{\tn{SVH}}(T)$, as a function of $T$ for different filling fractions. The SVH critical temperature, $T_c^{\tn{SVH}}$, is determined from the crossing point (black star) for different system sizes. }
\end{figure}

\begin{figure}[h!]
\includegraphics[width=170mm,scale=1]{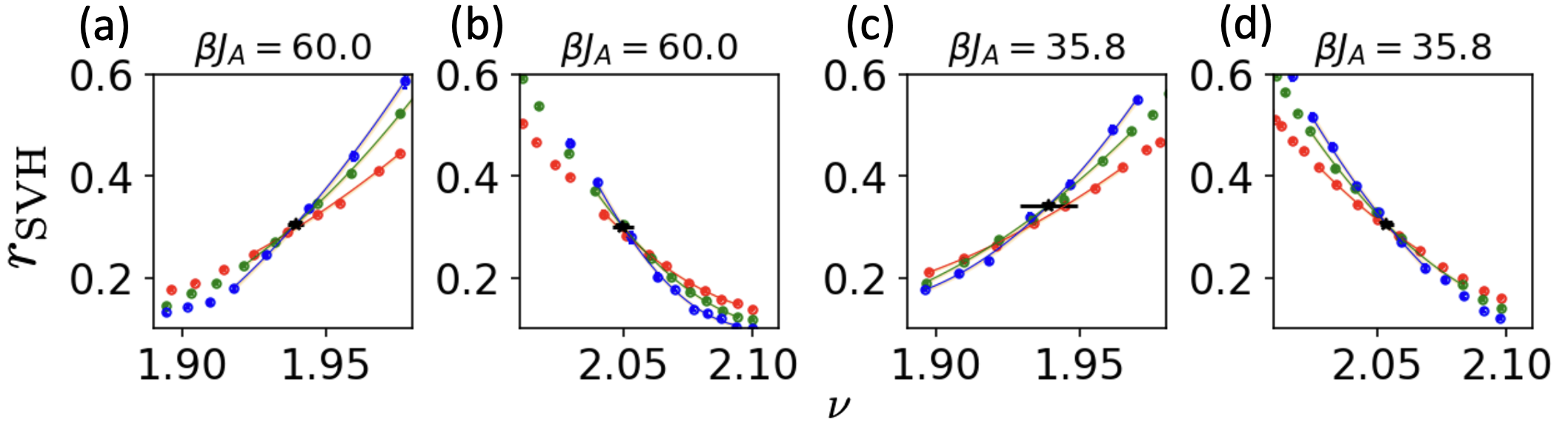}% Here is how to import EPS art
\caption{\label{nc_svh} RG-invariant correlation length, $r_{\tn{SVH}}$, as a function of filling fraction at two different temperatures. The solid lines are obtained from a least squares fit, with the yellow shaded region denoting the error generated by the fitting process. $\nu_c^{\tn{SVH}}$ is determined by the crossing point of the largest two system sizes. }
\end{figure}

\section{Determination of $J_\tn{H}^{\tn{c}1}$ and $J_\tn{H}^{\tn{c}2}$}\label{App::jhc}
We provide here additional details for determining the critical points $J_\tn{H}^{\tn{c}1}$ and $J_\tn{H}^{\tn{c}2}$ in the anisotropy-tuned transition between the SVH and SC phases. The determination is based on a polynomial fit of the scaling function in the finite-size scaling analysis. Specifically, we assume the RG-invariant correlation length satisfies the following scaling form \cite{fakher_scaling},
\beq\label{scaling_func1}
r_O(J_\tn{H}, L) \equiv \frac{\xi_O(J_\tn{H}, L)}{L} = f_\xi(J_{\tn{RG}}) + L^{-\omega} g_{\xi}(J_{\tn{RG}}),
\eeq
where $J_{\tn{RG}}$ is the RG-invariant tuning parameter defined as $J_{\tn{RG}} \equiv [(J_{\tn{H}}-J_{\tn{H}}^{\tn{c}})/J_{\tn{H}}^{\tn{c}}]L^{1/\nu_{\tn{RG}}}$. Here, $f_\xi(J_{\tn{RG}})$ is the scaling function that corresponds to the contribution from the relevant operator $O$ across the transition, and $g_\xi(J_{\tn{RG}})$ is the scaling function that captures the contribution from the leading irrelevant operator.

To determine the critical point $J_\tn{H}^c$, we use a polynomial fit for $f_\xi(J_{\tn{RG}})$ and $g_\xi(J_{\tn{RG}})$ up to cubic order, with the coefficients in the polynomial, $\omega$ and $J_\tn{H}^c$ as the fitting parameters to be determined by the least square fitting. The $\nu_{\tn{RG}}$ is fixed as an input during the fitting process \cite{Cardy_1996} as $\nu^{\tn{Ising}}_{\tn{RG}} = 0.630$ and $\nu^{\tn{XY}}_{\tn{RG}} = 0.672$ in 3D. The results for the fitting process are shown in Fig.~\ref{fss_jhc}a and Fig.~\ref{fss_jhc}b, for $r_{\tn{SC}}$ and $r_{\tn{SVH}}$, respectively. The colored data points with error-bar are the raw data obtained from QMC calculation, and the dashed lines denote the fitted curves $r_O^{\tn{fit}}$ for each system size. In the inset of each plot, we also show the same quantity as a function of $J_{\tn{RG}}$, where the black dots denote the fitting result. The color scheme is the same as the main text. The fitting results for $J_\tn{H}^c$ are shown in each plot.

In addition, we also use a different ansatz for the scaling function defined as
\beq\label{scaling_func2}
r_O(J_\tn{H}, L) \equiv \frac{\xi_O(J_\tn{H}, L)}{L} = f_\xi(J_{\tn{RG}}),
\eeq
where we ignore the contribution from the irrelevant operators. The associated fit results are shown in Fig.~\ref{fss_jhc}c and Fig.\ref{fss_jhc}d, for $r_{\tn{SC}}$ and $r_{\tn{SVH}}$, respectively. Thus, the $J_\tn{H}^c$ we obtain from the two fitting procedures are independent of the ansatz for scaling functions up to the fitting error.

\begin{figure}[h!]
\includegraphics[width=140mm,scale=1]{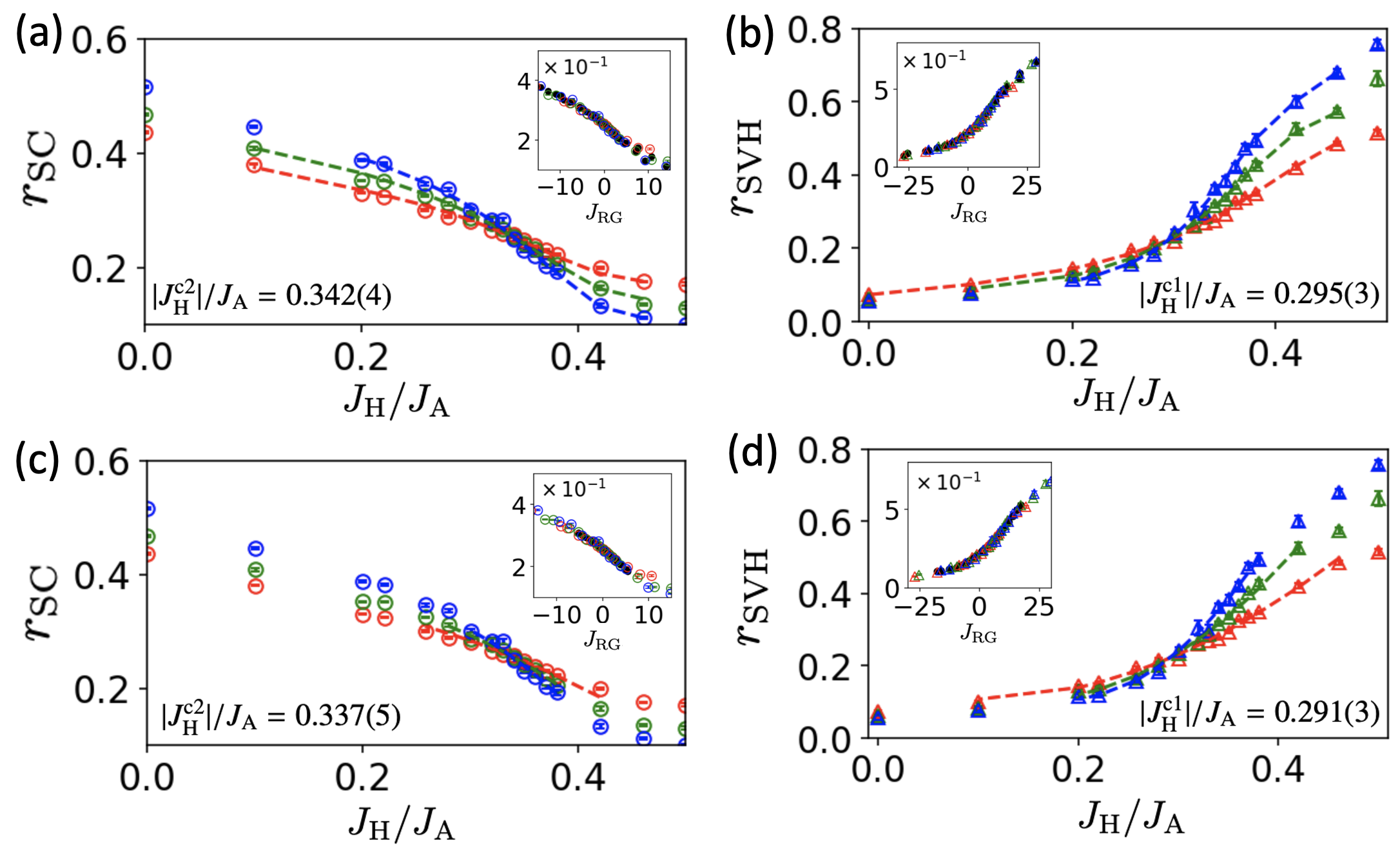}% Here is how to import EPS art
\caption{\label{fss_jhc} Finite-size scaling analysis of the RG-invariant correlation length. Data points with error-bar are obtained from the QMC calculation (color scheme is same as the main text). The dashed curves are obtained from a polynomial fitting of the respective scaling functions. (a) $r_{\tn{SC}}$, using the scaling function Eq.\ref{scaling_func1} (b) $r_{\tn{SVH}}$, using the scaling function Eq.\ref{scaling_func1} (c) $r_{\tn{SC}}$, using the scaling function Eq.\ref{scaling_func2} (d) $r_{\tn{SVH}}$, using the scaling function Eq.\ref{scaling_func2}. Insets show the plot of the same quantity as a function of RG-invariant tuning parameter  $J_{\tn{RG}}=[(J_{\tn{H}}-J_{\tn{H}}^{\tn{c}})/J_{\tn{H}}^{\tn{c}}]L^{1/\nu_{\tn{RG}}}$, where the colored data points are from QMC calculation and black dots are obtained from the polynomial fitting.}
\end{figure}

\section{Effective Non-Linear Sigma Model and Chemical potential tuned transition}\label{App::nlsm}
Here we provide additional details on the effective theory and the chemical potential tuned transition between the different phases. Recall the effective Hamiltonian in the low-energy Hilbert space, ${\cal{H}}_{\rm{eff}}$, introduced in Eqn.~\ref{proj_ham_nlr} in the main text,
\beq
H_{\tn{eff}} = \sum_{\r,\alpha} \bigg[G \bigg( U_\alpha L_{\alpha,\r}^2 + V_\alpha n_{\alpha, \r}^2 \bigg) + \mu L_{3, \r}\bigg] + \sum_{\r,\r';\alpha} F_{\r,\r'}\bigg[ U'_{\alpha}~L_{\alpha}(\r) L_{\alpha}(\r') +V'_{\alpha}~ n_{\alpha}(\r) n_{\alpha}(\r')  \bigg],
\eeq
where the symbols and coupling constants are as before. 

The operators $\{n_\alpha, L_\alpha\}$ defined in Eq.\ref{na_main} and Eq.\ref{La_main} satisfy the following commutation relation,
\begin{equation}\label{so4_comm}
\begin{aligned}
&
[n_\alpha, n_\beta] = i \varepsilon_{\alpha \beta\gamma} L_{\gamma}\\&
[L_\alpha, L_\beta] = i \varepsilon_{\alpha \beta\gamma} L_{\gamma}\\&
[L_\alpha, n_\beta] = i \varepsilon_{\alpha \beta\gamma} n_{\gamma}
\end{aligned}
\end{equation}
Let us demonstrate the physical meaning of the operators $\{n_\alpha, L_\alpha\}$ by making an explicit analogy with a bipartite anti-ferromagnetic spin system. Denoting the physical spin operators on a bipartite lattice indices by $\vec{S}^A$ and $\vec{S}^B$, we can define the uniform and staggered spin polarization as $\Tilde{L}_\alpha = S^A_\alpha + S^B_\alpha$ and $\Tilde{n}_\alpha = S^A_\alpha - S^B_\alpha$, respectively. It can be easily checked that $\{\tilde{n}_\alpha, \tilde{L}_\alpha\}$ satisfies the same commutation relation Eq.\ref{so4_comm}, which inspires us to map our effective Hamiltonian Eq.~\ref{proj_ham_nlr} to a pseudo-spin picture where $L_\alpha$ denotes the uniform pseudo-spin polarization and $n_\alpha$ denotes the staggered pseudo-spin polarization. 

Let us now discuss the chemical potential tuned transitions at a fixed $|J_\tn{H}|/J_\tn{A} = 0.5$ based on the above model; a schematic phase diagram in the pseudo-spin language is shown in Fig.\ref{mu_phase_diagram_supp}a. 

At $T=0$ and $\mu=0$, the system is in the N\'eel state due to the easy-axis anisotropy, leading to $\avg{n_3}\neq0$, which represents the incompressible SVH state. A finite $\mu_{c1}$ is required to dope a finite density of excess charge with $\avg{L_3}\neq0$ into the system. The compressiblity $\chi_c$ for the three lowest temperatures obtained from our quantum Monte-Carlo simulations are shown in Fig.~\ref{mu_phase_diagram_supp}b, c and d, respectively. 

Indeed, for $\mu_{c1}<\mu<\mu_{c2}$, the pseudo-spins are tilted slightly away from the perfect N\'eel arrangement, which denotes a finite density of extra charge $\avg{L_3}\neq0$ and a $\avg{n_3}\neq 0$ SVH order;  the in-plane component stays anti-aligned denoting an SC ordered phase co-existing with the SVH order. We expect the valley-singlet SC pair susceptibility $\chi_{\tn{vSC}}$ to be enhanced in the coexistent SVH-SC phase, as suggested by the remnant $\avg{L_{1,2}}\neq0$ in Fig.\ref{mu_phase_diagram_supp}a. This is clearly supported by the additional bump in the numerical data for $\chi_{\tn{vSC}}$ at the lowest temperature in Fig.~\ref{mu_phase_diagram_supp}e, when compared to the higher temperature data in Figs.~\ref{mu_phase_diagram_supp}f and g.

Increasing $\mu>\mu_{c2}$ flips the z-component, aligning the pseudospin completely with the external field; this indicates the disappearance of the SVH order, $\avg{n_3}= 0$, while the in-plane component denoting a pure spin-singlet SC stays anti-aligned, $\avg{n_{1,2}}= 0$.

\begin{figure}[h!]
\includegraphics[width=160mm,scale=1]{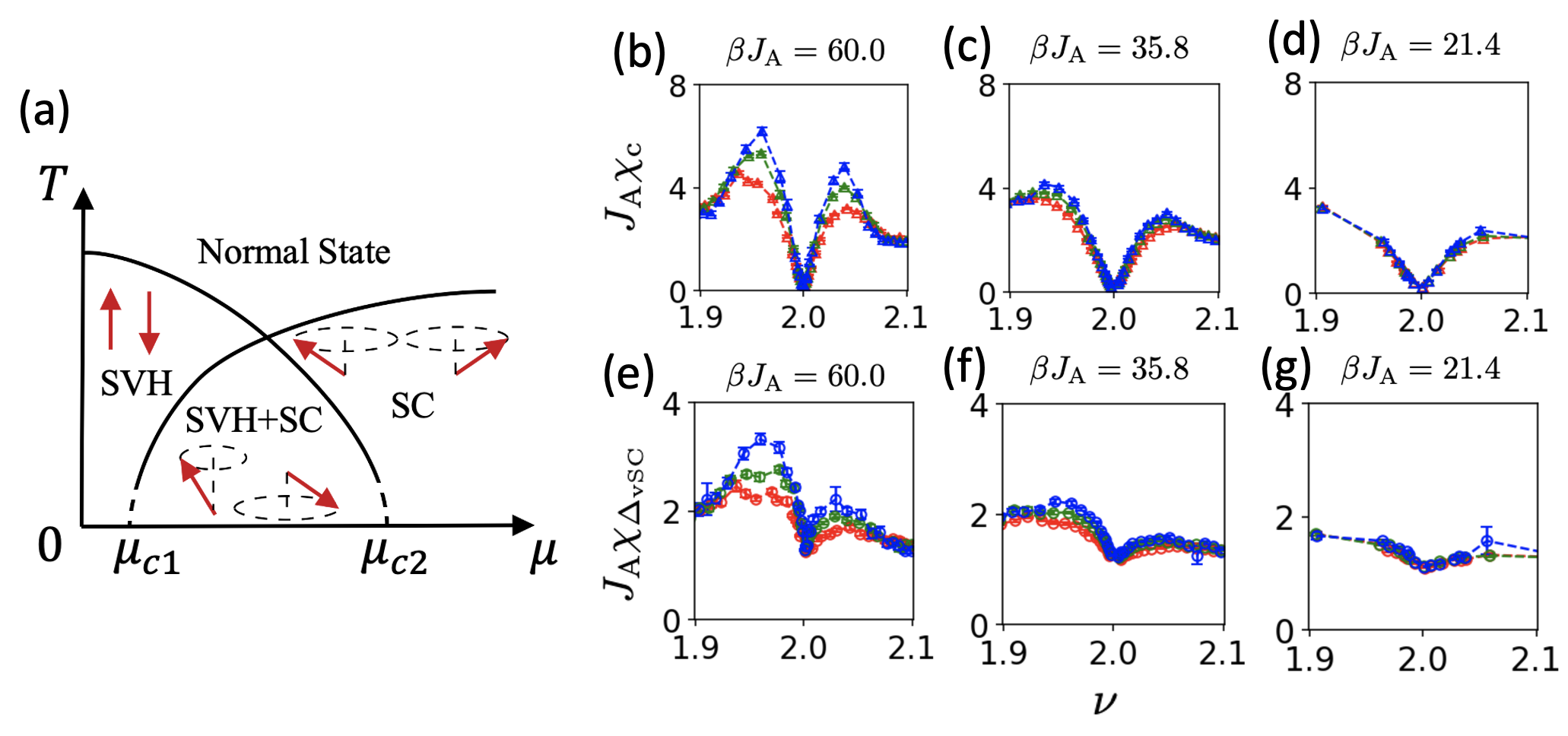}% Here is how to import EPS art
\caption{\label{mu_phase_diagram_supp} (a) Schematic $T-\mu$ phase-diagram in terms of the anti-ferromagnetic pseudo-spin effective model \cite{nelson-fisher1}. At $\nu=2.0$ and $\beta J_\tn{A} = 60.0$, the system is almost incompressible, and we plot the compressibility $\chi_c$ as a function of filling fraction $\nu$ at fixed temperature (b) $\beta J_\tn{A} = 60.0$, (c) $\beta J_\tn{A} = 35.8$, (d) $\beta J_\tn{A} = 21.4$, respectively. Valley-singlet SC pair susceptibility $\chi_{\tn{vSC}}$ is also plotted as a function of $\nu$ at fixed temperature (e) $\beta J_\tn{A} = 60.0$, (f) $\beta J_\tn{A} = 35.8$, (g) $\beta J_\tn{A} = 21.4$ . An enhancement of $\chi_{\tn{vSC}}$ is observed when doping away from $\nu=2.0$.}
\end{figure}

\end{widetext}